\documentclass[sigconf, nonacm]{acmart}

\usepackage{amsmath,amsfonts}
\usepackage{algorithm}
\usepackage{algpseudocode}
\usepackage{graphicx}
\usepackage{booktabs}
\usepackage{textcomp}
\usepackage{xcolor}
\usepackage{todonotes}
\usepackage{hyperref}
\usepackage{enumitem}
\usepackage{float}
\newcommand{\CommentAbove}[1]{\Statex \(\triangleright\) #1}
\AtBeginDocument{%
  }
\newcommand{\dynres}{\textsf{DynRes}}

\setcopyright{none}
\author{Keerthi Gaddameedi}
\orcid{0009-0003-0658-3884}
\affiliation{%
  \institution{Technical University of Munich}
  \city{Munich}
  \country{Germany}}
\email{keerthi.gaddameedi@tum.de}

\author{Dominik Huber}
\orcid{0000-0001-9696-9382}
\affiliation{%
  \institution{Technical University of Munich}
  \city{Munich}
  \country{Germany}}
\email{domi.huber@tum.de}

\author{Martin Schreiber}
\orcid{0000-0002-2390-6716}
\affiliation{%
  \institution{Université Grenoble Alpes}
  \city{Grenoble}
  \country{France}}
\email{martin.schreiber@univ-grenoble-alpes.fr}

\author{Hans-Joachim Bungartz}
\orcid{0000-0002-0171-0712}
\affiliation{%
  \institution{Technical University of Munich}
  \city{Munich}
  \country{Germany}}
\email{bungartz@cit.tum.de}

\author{Valentina Sch\"uller}
\orcid{0000-0002-4579-4217}
\affiliation{%
  \institution{Lund University}
  \city{Lund}
  \country{Sweden}}
\email{valentina.schuller@math.lu.se}

\author{Tobias Neckel}
\orcid{0000-0002-3442-7171}
\affiliation{%
  \institution{Technical University of Munich}
  \city{Munich}
  \country{Germany}}
\email{neckel@cit.tum.de}

\author{Martin Schulz}
\orcid{0000-0001-9013-435X}
\affiliation{%
  \institution{Technical University of Munich}
  \city{Munich}
  \country{Germany}}
\email{schulzm@in.tum.de}
\begin{document}

\title{Adaptive Parallel-in-Time Integration with Dynamic Resource Management}


\begin{abstract}
As computational resources continue to grow, the strong-scaling limitations of spatial parallelism motivate the pursuit of additional concurrency in the temporal dimension, 
particularly for applications with hard time constraints, such as weather and climate simulations. The Parallel Full Approximation Scheme in Space and Time (PFASST) is a parallel-in-time method 
based on Spectral Deferred Corrections (SDC).
It computes multiple timesteps concurrently by coupling fine- and coarse-grid SDC sweeps using multigrid Full Approximation Scheme (FAS) corrections.
However, PFASST's convergence is often problem-dependent, demanding a variable number of 
parallel timesteps and, hence, computing resources at different times throughout the simulation.
Dynamic Resource Management (DRM) provides a remedy for this challenge by enabling 
the adaptive adjustment of computational resources and algorithmic parameters at runtime.
In this work, we present our novel approach to extending PFASST with DRM, which enables (a) dynamic adaptation of computing resources,
(b) adaptive selection of the number of PFASST iterations based on local convergence behavior, and
(c) coupling of these two adaptations into a single resizing strategy.
With this approach, we demonstrate for the first time that optimal configurations can be identified in real time for each application, rather than relying on static allocation.
Furthermore, we show that convergence-informed tuning of PFASST improves resource utilization and convergence efficiency.
\end{abstract}
\settopmatter{printacmref=false}
\maketitle
\pagestyle{plain}
\section{Introduction}\label{sec:intro}
Historically, most optimization efforts for numerical methods have focused on domain decomposition and spatial parallelism~\cite{domdecsurvey}.
However, as the benefits of spatial parallelism begin to plateau and the availability of computational resources exceeds what can effectively 
be utilized for spatial parallelism alone, there has been a growing interest in developing and improving methods that also
incorporate parallelism in the temporal dimension. 
Parallel Full Approximation Scheme in Space and Time (PFASST)~\cite{pfasst} is a parallel-in-time method that first computes an
inexpensive coarse approximation serially across a block of timesteps, then iteratively refines this approximation in parallel
using more expensive fine-level corrections.
Although PFASST enables efficient time-parallel computations, its accuracy can diminish as the number of parallel timesteps increases. 
A rigorous convergence analysis confirms this degradation for a fixed total time interval, as increasing the number of parallel timesteps 
causes the step size to shrink accordingly~\cite{convTheory}. For the complementary scenario, extending the covered interval while keeping 
the step size fixed, which is relevant when varying the block size, no analogous theoretical bound has been established yet. 
However, empirical studies have demonstrated this degradation in precisely such settings~\cite{spacetimeheat, pfasstsh}.
To address this, we propose leveraging Dynamic Resource Management (DRM)~\cite{drmheat}, the runtime reallocation of computational resources
to a PFASST block to enable convergence-driven adaptivity of algorithmic parameters such as the block size, i.e., the number of parallel
timesteps.
By adjusting these parameters at runtime based on observed convergence behavior, rather than fixing
them in advance, DRM allows PFASST to trade off the degree of temporal parallelism against the accuracy penalty it incurs,
targeting an optimal balance between convergence and performance.
We demonstrate the resulting adaptive PFASST algorithm on a solver for the barotropic equations, a simplified but
representative model for oceanic and atmospheric dynamical cores, which makes it a 
well-suited testbed for exploiting additional parallelism in the temporal dimension~\cite{sdcsphere, mlsdcsh, pfasstsh}.

\subsection{Related Work}
Over the past few decades, a wide range of numerical methods have been developed to improve the accuracy and efficiency of weather and climate
modeling~\cite{williamson2007}. Dynamical cores are especially challenging because of the wide range of temporal scales involved: fast
acoustic and gravity waves require small timesteps for explicit stability under the CFL condition, while slower large-scale dynamics make
fully implicit schemes expensive. High-order methods such as Spectral Deferred Corrections (SDC)~\cite{sdc} and Multi-Level SDC (MLSDC)~\cite{mlsdc} enable larger
timesteps for the Shallow Water Equations (SWE) on the sphere~\cite{sdcsphere, mlsdcsh}.

A further challenge is achieving computational efficiency at exascale, motivating growing interest in parallel-in-time (PinT) methods such
as Parareal~\cite{parareal}, Multigrid Reduction in Time (MGRIT)~\cite{mgrit}, Rational Approximation of Exponential Integrators (REXI)
\cite{rexi, schreiberloft2019}, and PFASST~\cite{pfasst}. These have been combined with various spatial discretizations for the SWE,
including finite differences with Parareal~\cite{pararealswe}, and spherical harmonics with exponential-integration methods~\cite{rexish},
ParaDiag~\cite{paradiag}, MGRIT~\cite{mgritsh}, and PFASST~\cite{pfasstsh}.
Parareal-type methods are known to struggle with pure advec\-tion-dominated, hyperbolic, wave-dominated
problems~\cite{ruprecht2018}. This was addressed by developing optimized coarse-grid operators for MGRIT and Parareal that
track characteristic curves rather than relying on standard rediscretization, improving convergence on linear
advection problems~\cite{desterck2021}. This was later extended with a local Fourier analysis convergence theory
for two-level MGRIT on advection-dominated PDEs, showing that poor convergence stems from inadequate treatment of
certain smooth Fourier modes~\cite{desterck2025}. However, these approaches redesign the coarse-grid operator itself---a structural choice made once, in advance,
independent of how the solver behaves at runtime. A complementary strategy is to instead let the algorithm's runtime
parameters such as timestep size, iteration count, and block size adapt dynamically to observed convergence, rather than fixing them a priori.
Adaptive SDC~\cite{adaptivesdc}, for instance, adjusts timestep size and iteration count from local convergence estimates, matching
state-of-the-art adaptive Runge-Kutta efficiency. Adaptive Parareal~\cite{adaptiveparareal} improves fine-solver accuracy over iterations,
with the coarse-solver cost remaining as the main obstacle to full scalability. Adaptive spatial coarsening has similarly improved MGRIT
convergence where wave speeds are locally near zero~\cite{adaptivemgrit}, and an adaptive Parareal variant for molecular dynamics
\cite{adaptivepararealmd} adapts both slab size and iteration count to outperform the classical algorithm. Each method, however, also has
regimes where its benefits diminish: for example, adaptive SDC under loose tolerances~\cite{adaptivesdc}, and adaptive Parareal under high coarse-solver cost
\cite{adaptiveparareal}. For PFASST specifically, however, this performance-accuracy balance has not yet been demonstrated, despite PFASST-SH
\cite{pfasstsh} showing that accuracy degrades directly with increasing parallelism.

Dynamic resource adaptivity offers a complementary axis to this algorithmic adaptivity. TALP, a profiling tool from the DLB library, has been
used to adapt resources in the CFD code Alya~\cite{Alya} via a communication-efficiency metric, reaching target efficiency within a few steps
regardless of initial allocation~\cite{houzeaux2022}. The DMR middleware applies the same metric to GROMACS~\cite{gromacs}, again
outperforming static allocation~\cite{sandas2026}. Closer to our work, \dynres{} software has been combined with PFASST to show the feasibility of
dynamically resizing a block~\cite{drmheat}. These approaches, however, rely on performance-driven signals rather than the solver's own
numerical behavior. This distinction matters for PFASST, since PFASST-SH~\cite{pfasstsh} shows that a purely performance-driven signal is
insufficient once parallelism starts degrading accuracy.

\subsection{Contributions}
To enable convergence-informed resource adaptivity in PFASST, avoiding the need to search for an optimal configuration in advance~\cite{pfasster}, we make the following main contributions:
\begin{itemize}[leftmargin=3.5mm]
    \item We extend PFASST with \dynres{}, enabling a PFASST block to dynamically 
    grow or shrink its process allocation at runtime in response to convergence-based 
    triggers, rather than relying on a predetermined, fixed allocation.
    \item We investigate the impact of two key parameters---block size and iteration budget---that significantly influence both the convergence and performance of PFASST. 
    Building on this insight, we design and implement two concrete DRM-driven resizing strategies: \textsc{parallel\_steps\_resizer}, which adapts block size, 
    and \textsc{iters\_resizer}, which dynamically adjusts both block size and the iteration budget relative to it.
    \item We evaluate both strategies on two PDE benchmarks, quantifying and analyzing effects on parallelism adaptation, 
    performance, convergence, and resource utilization.
\end{itemize}

\section{Parallel Full Approximation Scheme in Space and Time}
PFASST~\cite{pfasst} is an iterative, multi-level time integration method for Initial Value Problems (IVPs).
Unlike conventional time integration methods, PFASST computes multiple timesteps in parallel by combining an arbitrarily high-order time
 integration method---Spectral Deferred Corrections (SDC)~\cite{sdc}---with a multigrid correction strategy called the Full Approximation Scheme (FAS)~\cite{fas}.
 The algorithm begins by computing a low-order SDC approximation for a block of timesteps in serial, then iteratively refines these 
 approximations in parallel using a finer grid. Over successive iterations, higher-order approximations are propagated within the block, 
 progressively improving the solution. In the subsequent sections, we first describe SDC and its multi-level variant
 Multi-Level Spectral Deferred Corrections (MLSDC)~\cite{mlsdc}, and then explain how these components are combined to form the PFASST algorithm.
Readers less familiar with applied mathematics may skip directly to Section~\ref{sec:drm} on Dynamic Resource Management.

\subsection{Spectral Deferred Corrections}
SDC~\cite{sdc} is a family of iterative methods, developed as a high-order extension of Defect Correction~\cite{defectcorrect} and Deferred Correction
methods~\cite{IteratedDC,deferredcorrect}. SDC solves the collocation equation obtained from a high-order quadrature approximation of the Picard integral formulation
by performing a sequence of low-order correction iterations that progressively increase the order of accuracy.

We outline the steps of SDC using an IVP given by
\begin{equation}
\frac{dU(t)}{dt} = F(U(t),t), \qquad U(0) = U_0,
\label{eq:ivp}
\end{equation}
where $t \in [0,T]$, $U_0,\,U(t) \in \mathbb{C}^N$, and
$F : \mathbb{C}^N \times \mathbb{R} \rightarrow \mathbb{C}^N$.
The Picard formulation obtained by integrating \eqref{eq:ivp} for a given time interval $[0,t]$ is given by:
\begin{equation}
U(t) = U_0 + \int_{0}^{t} F(U(\sigma),\sigma) d\sigma.
\label{eq:picard}
\end{equation}
SDC computes solutions on one time interval at a time. A time interval $[t_n,t_{n+1}]$ is first split into substeps which
correspond to nodes of a quadrature rule. We use Gauss--Lobatto quadrature here and therefore define $M+1$ nodes on the
given interval as $t_n = t_0 < t_1 < \cdots < t_M = t_{n+1}$, with $\Delta t \equiv t_{n+1} - t_n$, the length of the
full interval, and $\Delta t_m \equiv t_{m+1} - t_m$, the length of substep $m$.
We can approximate $F(U(\sigma),\sigma)$ at each node in the interval using Lagrange polynomial interpolation.
%

Discretizing \eqref{eq:picard} and replacing $F(U(\sigma),\sigma)$ with its $j$-th Lagrange polynomial interpolant $L_j(\sigma)$~\cite{hairer} at node $m$ gives us
\begin{equation}
U_m = U_0 + \sum_{j=0}^{M} F(U_j,t_j) \int_{t_0}^{t_m} L_j(\sigma)\, d\sigma,
\label{eq:collocation_integral}
\end{equation}
where $m = 0,\ldots,M$.

Given the quadrature weights $q_{m,j}$, representing the weight at node $m$ due to the
contribution at node $j$,
we obtain the collocation equation:
\begin{equation}
U_m = U_0 + \Delta t \sum_{j=0}^{M} q_{m,j}\,F(U_j,t_j).
\label{eq:collocation}
\end{equation}
Then, we compute an initial low-order approximation using, say, the Implicit Euler method.
The correction sweep uses local quadrature weights $s_{m,j}$, which integrate over a single
substep $[t_m,t_{m+1}]$ rather than cumulatively from $t_0$ as $q_{m,j}$ does,
giving
\begin{equation}
s_{m,j} = q_{m+1,j} - q_{m,j}, \qquad m=0,\ldots,M-1,\ j=0,\ldots,M.
\label{eq:s_from_q}
\end{equation}
Then the correction equation, also referred to as a sweep, is given by:
\begin{equation}
\begin{aligned}
U_{m+1}^{k+1}
&= U_m^{k+1} + \Delta t_m
\left[
F(U_{m+1}^{k+1},t_{m+1})
-
F(U_{m+1}^{k},t_{m+1})
\right] \\
&\quad + \Delta t
\sum_{j=0}^{M}
s_{m,j}\,
F(U_j^{k},t_j),
\end{aligned}
\label{eq:sdc_implicit_sweep}
\end{equation}
for $m=0,\ldots,M-1$, where $k$ denotes the iteration.

\subsection{Multi-Level Spectral Deferred Corrections}\label{sec:mlsdc}
In MLSDC~\cite{mlsdc}, the SDC sweep from \eqref{eq:sdc_implicit_sweep} is performed across a hierarchy of levels using FAS as a correction scheme. 
The objective is to reduce the number of sweeps required on the fine discretization by using sweeps on the coarse discretization to compute corrections 
that enhance the solution on the finer grid.
We restrict the following derivation of the algorithm to two levels, as this is the configuration for the experiments of this paper.

We denote the fine level as $\ell=f$, coarse level as $\ell=c$, and denote the restriction operator (fine to coarse) as $\mathcal{R}$ and the interpolation operator
 (coarse to fine) as $\mathcal{I}$. 

We begin by computing an SDC sweep on the fine level $U_{m+1}^{f,k+1}$, similar to \eqref{eq:sdc_implicit_sweep}.
The restriction operator $\mathcal{R}$ is defined as
\begin{equation}
\begin{aligned}
\vec U^{c,k} &= \mathcal{R}\,\vec U^{f,k+1}, \\
\text{with} \quad \vec U^{\ell} &= \left(U_0^{\ell},\,U_1^{\ell},\,\dots,\,U_{M_\ell}^{\ell}\right) \in \mathbb{C}^{(M_\ell+1)N_\ell}
\end{aligned}
\label{eq:restrictoperator}
\end{equation}
for $\ell \in \{f,c\}$, where $\mathcal{R} \in \mathbb{C}^{(M_c+1)N_c \times (M_f+1)N_f}$ and $N_\ell$ is used to represent the spatial resolution.

The residual at node $m$ on level $\ell$ is the difference between two consecutive sweeps; stacking these node-wise
residuals as in \eqref{eq:restrictoperator} gives the residual vector
\begin{equation}
\begin{aligned}
\mathrm{Res}_m^{\ell,k} &= U_0^{\ell} + \Delta t \sum_{j=0}^{M_\ell} q_{m,j}^{\ell}\, F^{\ell}\big(U_j^{\ell,k},t_j\big) - U_m^{\ell,k}, \\
\text{with} \quad \overrightarrow{\mathrm{Res}}^{\,\ell,k} &= \left(\mathrm{Res}_0^{\ell,k},\,\mathrm{Res}_1^{\ell,k},\,\dots,\,\mathrm{Res}_{M_\ell}^{\ell,k}\right) \in \mathbb{C}^{(M_\ell+1)N_\ell},
\end{aligned}
\label{eq:mlsdc_residual_vec}
\end{equation}
for $m=0,\ldots,M_\ell$ and $\ell\in\{f,c\}$.

The FAS correction is then computed as the difference between the restricted residual on the fine level and the coarse-level residual:
\begin{equation}
\vec\tau^{\,c,k} = \mathcal{R}\,\overrightarrow{\mathrm{Res}}^{\,f,k+1} - \overrightarrow{\mathrm{Res}}^{\,c}\big(\vec U^{c,k}\big).
\label{eq:mlsdc_tau}
\end{equation}

An SDC sweep is then computed on the coarse level using $\vec\tau^{c,k}$,
\begin{equation}
\begin{aligned}
U_{m+1}^{c,k+1} &= U_m^{c,k+1} + \Delta t_m \left[F^c(U_{m+1}^{c,k+1},t_{m+1}) - F^c(U_{m+1}^{c,k},t_{m+1})\right] \\
&\quad + \Delta t \sum_{j=0}^{M_c} s_{m,j}^c\, F^c(U_j^{c,k},t_j) + \left(\tau_{m+1}^{c,k} - \tau_m^{c,k}\right),
\end{aligned}
\label{eq:mlsdc_coarse_sweep}
\end{equation}
for $m=0,\ldots,M_c-1$.

The resulting coarse-level correction $\vec U^{c,k+1} - \vec U^{c,k}$ is interpolated to the fine level using $\mathcal{I}$ and
added to the fine-level solution,
\begin{equation}
\vec U^{f,k+1} \;\leftarrow\; \vec U^{f,k+1} + \mathcal{I}\,\left(\vec U^{c,k+1} - \vec U^{c,k}\right),
\label{eq:mlsdc_interpolate}
\end{equation}
where $\mathcal{I} \in \mathbb{C}^{(M_f+1)N_f \times (M_c+1)N_c}$.

\subsection{PFASST Algorithm}

In PFASST~\cite{pfasst}, $N_{ts}$ timesteps are computed simultaneously, one per processor $P_n$ ($n=0,\ldots,N_{ts}-1$), using a serial coarse propagator to iteratively correct a fine propagator.
Extending the per-node, per-level notation from Section~\ref{sec:mlsdc} to this multi-processor setting, we write $U_{n,m}^{\ell,k}$ for the solution at quadrature node $m$ on level $\ell$ after iteration $k$, for the timestep assigned to processor $P_n$, and $\vec U_n^{\ell,k}$ for the corresponding vector stacked over all nodes $m$.
The algorithm can be divided into four phases, as also sketched out in Alg.~\ref{alg:pfasst_iteration}:
\begin{itemize}[leftmargin=3.5mm]
\item \textbf{Prediction Phase:} \textit{\textbf{Serial}} coarse sweeps are performed to propagate the initial value across all $N_{ts}$ timesteps in the block, which remains inexpensive since coarse 
sweeps operate on a much coarser grid than fine sweeps.
\item \textbf{Fine Sweep and FAS Correction Phase:} A fine SDC sweep is performed in \textit{\textbf{parallel}} on each timestep of the block, using the initial values obtained from the prediction phase.
FAS corrections are also computed in \textit{\textbf{parallel}} by calculating the difference between the restricted fine-level residuals and the coarse-level residuals at each timestep. The fine sweep is
the most computationally expensive step, and computing many timesteps in parallel is what yields PFASST its significant speedup over traditional serial methods.
\item \textbf{Coarse Sweep Phase:} The next coarse sweep is performed \textit{\textbf{serially}}, incorporating the FAS correction, and the value at the last 
quadrature node of each timestep is communicated to the subsequent processor for its coarse sweep.
\item \textbf{Interpolation Phase:} The coarse-level corrections are interpolated to the fine level in \textit{\textbf{parallel}}, producing the initial condition
used by the fine sweep at the start of the next iteration.
\end{itemize}
The last three phases are repeated iteratively until the residual at the final quadrature node of
each timestep, maximized across all $N_{ts}$ parallel timesteps, falls below a predetermined
tolerance or the maximum number of iterations ($K^{\max}$) is reached.
\vspace{-0.75em}
\begin{algorithm}
\caption{PFASST algorithm for $N_{ts}$ parallel timesteps}
\label{alg:pfasst_iteration}
\begin{algorithmic}[1]
\For{$n = 0,\dots,N_{ts}-1$ \textbf{in serial}} \Comment{Prediction}
    \State $U_{n,0}^{c,0} \gets$ initial condition $\vec U_0$ if $n=0$, else last coarse node from $P_{n-1}$
    \State $\vec U_n^{c,0} \gets$ coarse SDC sweep on $\vec U_n^{c,0}$
    \State send $U_{n,M_c}^{c,0}$ to $P_{n+1}$ if $n<N_{ts}-1$
    \State $\vec U_n^{f,0} \gets \mathcal{I}\,\vec U_n^{c,0}$
\EndFor
\For{$k = 0, 1, 2, \dots K^{\max}$}
    \For{$n = 0,\dots,N_{ts}-1$ \textbf{in parallel}} \Comment{Fine sweep \& FAS correction}
        \State $\vec U_n^{f,k+1} \gets$ fine SDC sweep on $\vec U_n^{f,k}$
        \State $\vec U_n^{c,k} \gets \mathcal{R}\,\vec U_n^{f,k+1}$; \quad $\vec\tau_n^{\,k} \gets \mathcal{R}\,\overrightarrow{\mathrm{Res}}_n^{\,f,k+1} - \overrightarrow{\mathrm{Res}}_n^{\,c,k}$
    \EndFor
    \If{$\displaystyle \max_{0\le n<N_{ts}} \left\|\mathrm{Res}_{n,M_f}^{k+1}\right\| < \text{tolerance}$}
        \State \textbf{break}
    \EndIf
    \For{$n = 0,\dots,N_{ts}-1$ \textbf{in serial}} \Comment{Coarse sweep}
        \State $U_{n,0}^{c,k} \gets$ last coarse node from $P_{n-1}$ if $n>0$
        \State $\vec U_n^{c,k+1} \gets$ coarse SDC sweep on $\vec U_n^{c,k}$ using $\vec\tau_n^{\,k}$
        \State send $U_{n,M_c}^{c,k+1}$ to $P_{n+1}$ if $n<N_{ts}-1$
    \EndFor
    \For{$n = 0,\dots,N_{ts}-1$ \textbf{in parallel}} \Comment{Interpolation}
        \State $\vec U_n^{f,k+1} \gets \vec U_n^{f,k+1} + \mathcal{I}\,\left(\vec U_n^{c,k+1} - \vec U_n^{c,k}\right)$
    \EndFor
\EndFor
\end{algorithmic}
\end{algorithm}
\vspace{-0.75em}
Alg.~\ref{alg:pfasst_iteration} advances a single block of $N_{ts}$ timesteps. The simulation is composed of a sequence of such blocks, with the final state of a block providing the initial condition for the next.
This formulation serves as the foundation for the resource-adaptive algorithm we propose in this work. 
\section{Dynamic Resource Management} \label{sec:drm}
The static execution model of current HPC systems restricts jobs to the resources allocated at startup. The lack of system-level mechanisms for dynamically changing these allocations during execution limits the exploration and development of applications and scheduling algorithms that can adapt their resource usage at runtime. Establishing Dynamic Resource Management (DRM) on HPC systems, therefore, requires the co-design of system software and applications to overcome these static resource constraints. Early work has shown that DRM can improve key system metrics, including throughput, job turnaround time, and energy efficiency~\cite{iserte20,tarraf24,drmheat}, motivating its adoption in production systems.

However, adapting HPC applications and libraries to support DRM remains challenging. Runtime resource changes require applications to interact with the system scheduler, dynamically manage their processes, and redistribute application data according to their specific data structures and requirements.
To address these challenges, \textit{Dynamic Processes with PSets (DPP)}~\cite{huber24} provides a generic abstraction based on Process Sets (PSets) and Process Set Operations (PSetOps). PSets are unique labels for subsets of application processes, enabling fine-grained references to processes and their associated resources. By specifying PSetOps over lists of input PSets, applications can communicate to the resource manager the reconfigurations they support for individual application components. The resource manager can then select a concrete reconfiguration according to its optimization objective and represent the result as a list of output PSets. A resource reconfiguration can therefore be expressed as a subgraph
$$
[P^0_{in}, \cdots, P^N_{in}]
\xrightarrow{PSetOp}
[P^0_{out}, \cdots, P^M_{out}],
$$
where a PSetOp is a directed hyperedge connecting vertices representing PSets. The dynamic execution graph of an application is the union of these subgraphs, providing a general representation of the application's dynamic behavior.

In our work, for example, we use a REPLACE operation to increase and/or decrease the number of processes and their associated resources. The REPLACE operation is defined as

$$
[P_{in}]
\xrightarrow{REPLACE}
[P_{sub}, P_{add}, P_{replace}],
$$

where $P_{add}$ contains processes added by the operation, $P_{sub}$ contains processes removed by the operation, and
$P_{replace} = (P_{in} \setminus P_{sub}) \cup P_{add}.$

The DPP design has been implemented in the \textit{DynRes software stack}~\cite{dynres}, enabling DRM for different applications and programming models~\cite{drmheat,ju25,posner25}. In this work, we add Fortran bindings to an Open MPI-based layer of the \dynres software stack that extends the MPI Sessions interface with DPP. This extended interface allows us to specify the dynamic execution structure of PFASST and associate additional data with PSets and PSetOps to inform the resource manager's optimization decisions.

\section{Adaptive Parallel Full Approximation Scheme in Space and Time}
As established in Section~\ref{sec:intro}, fixing PFASST's parameters in advance cannot balance its accuracy-parallelism
trade-off without costly, problem-specific tuning. Simply
adapting algorithmic parameters, however, may improve convergence while degrading performance; we therefore
turn to Dynamic Resource Management (DRM) to jointly tune PFASST's algorithmic parameters and its computational
resources at runtime, using the solver's own convergence behavior as feedback.
\subsection{Limitations of PFASST} \label{sec:limitations}
PFASST's convergence has been documented across a wide range of applications and is known to be problem-dependent~\cite{50Years, pintApps}.
For advection-dominated or highly oscillatory problems, the coarse level can represent the underlying dynamics less accurately, 
which can deteriorate convergence of PFASST and increase the number of iterations required to reach a prescribed tolerance~\cite{convTheory}.
Such an effect in the Parareal case has been attributed to phase errors arising from differences between the coarse and fine propagators~\cite{ruprecht2018}.
The cost of convergence can depend on the number of parallel timesteps in a PFASST block, $N_{ts}$, since the PFASST iteration operates on a composite collocation problem coupling the individual timesteps.
Parareal-type coarse/fine correction schemes are known to converge exactly within $N_{ts}$ iterations for linear problems~\cite{50Years},
because each iteration there resolves every timestep with the exact fine propagator. PFASST's fine sweep \eqref{eq:sdc_implicit_sweep},
by contrast, performs only a limited number of SDC sweeps per timestep rather than solving the collocation problem exactly. This cheaper,
approximate correction is what makes each PFASST iteration affordable, but it also removes the guarantee that convergence is reached after
$N_{ts}$ iterations~\cite{unifiedpint}, so the number of iterations required cannot be bounded a priori and must instead be determined empirically.

An empirical study of PFASST applied to the rotating Shallow Water Equations has shown that, with a fixed number of iterations, increasing the block size (i.e., the degree of parallelism) degrades accuracy and, 
in some cases, leads to instability~\cite{pfasstsh}. The same study has also shown that the number of iterations required for convergence varies across different benchmarks. 
As a result, configuring parameters such as the number of iterations and parallel timesteps in advance can be detrimental to either convergence or performance.
These problem-dependent limits are exactly what our adaptive framework, described next, is designed to navigate at
runtime with the help of \dynres.

\subsection{Adaptive Framework}\label{sec:adaptiveframework}
We focus on two PFASST parameters---block size and number of PFASST iterations---that directly influence convergence and time-to-solution, but whose optimal configuration
is not known in advance. To assess convergence at runtime, we augment the fixed number of PFASST iterations with a residual-based stopping criterion, so that
Alg.~\ref{alg:pfasst_iteration} terminates once either the maximum norm of the residual falls below a specified tolerance or $K^{\max}$ iterations have been performed.
We invoke the resizing strategy once per block $b$, and the resulting adaptive scheme is given in Alg.~\ref{alg:adaptive_template}. Here, $\mathcal{H}_b$
denotes the convergence history of block $b$. \textsc{ResizeStrategy} returns a resize delta $\Delta$, in units of $P_{\text{node}}$ processors (the
number of cores available on one compute node), which is to be
applied to the block size for block $b+1$, together with an updated iteration ceiling $\tilde{K}^{\max}$. When \textsc{ResizeStrategy} is
\textsc{parallel\_steps\_resizer} (Alg.~\ref{alg:parallel_steps_resizer}), $\tilde{K}^{\max} = K^{\max}$ is simply passed through unchanged.
\vspace{-0.5em}
\begin{algorithm}
\caption{Adaptive PFASST}
\label{alg:adaptive_template}
\begin{algorithmic}[1]
\For{each block $b = 0, 1, 2, \dots$}
    \State $\mathcal{H}_b \gets$ Run Alg.~\ref{alg:pfasst_iteration} for block $b$
    \State $(\Delta, \tilde{K}^{\max}) \gets \textsc{ResizeStrategy}(N_{ts}, K^{\max}, \mathcal{H}_b)$
    \State $N_{ts} \gets N_{ts} + \Delta$
    \State $K^{\max} \gets \tilde{K}^{\max}$
\EndFor
\end{algorithmic}
\end{algorithm}
\vspace{-0.2em}
\subsection{Block Size Adaptation}
The first resizing strategy adjusts the number of parallel timesteps in a PFASST block based on the convergence speed of the current block.
To quantify this, we define an average convergence rate, $\rho_{avg}$, as shown in line~\ref{line:rho_avg} of 
Alg.~\ref{alg:parallel_steps_resizer}. This rate is calculated as the geometric mean of the per-iteration contraction of the residual over the 
$k$ iterations required for the block to converge, when convergence is achieved. If convergence is reached at the maximum allowed number of iterations, 
or not achieved at all, $k$ is equal to $K^{\max}$. 
Then, in line~\ref{line:iterratio}, we compute $\eta_k$, the fraction of the iteration budget $K^{\max}$ consumed before convergence.
Neither metric alone is sufficient to reliably classify a block's convergence speed. A block may contract quickly per iteration yet still consume most of its iteration budget if it starts from a poor initial residual. 
Therefore, the resizing decision requires both a low $\rho_{avg}$ and a low $\eta_k$ to classify a block as having converged quickly. Furthermore, we label a block as converged in line~\ref{line:converged} if the residual of the last iteration of the block
goes below a tolerance $\varepsilon_{\text{tol}}$, which is defined as $10^{-8}$ throughout this work.
$\rho_{\text{fast}} = 0.8$ and $\rho_{\text{slow}} = 0.3$ are threshold values against which $\rho_{avg}$ is compared to classify a block's convergence rate as fast or slow,
and $\alpha_{\text{fast}} = 0.5$ is a threshold on $\eta_k$, allowing a block that has consumed up to half its iteration budget to be classified as fast-converging.
To prevent a single noisy or borderline block from triggering an immediate resize, each grow or shrink decision is 
filtered through a \textsc{Hysteresis} mechanism, as described in Alg.~\ref{alg:hysteresis}. 
Given a proposed change $\delta$, a per-direction counter $c$, and a threshold $H$, 
the counter $c$ is incremented with each consecutive trigger in the same direction. 
Only when $c$ reaches $H$ does the algorithm apply the resize (returning $\delta$) and reset $c$ to $0$.
Otherwise, it returns $0$ and keeps the block size unchanged.
\vspace{-0.75em}
\begin{algorithm}
\caption{\textsc{Hysteresis}({$\delta$, $c$, $H$})}
\label{alg:hysteresis}
\begin{algorithmic}[1]
\State $c \gets c + 1$
\If{$c \geq H$}
    \State $c \gets 0$
    \State \Return $\delta$
\EndIf
\State \Return $0$
\end{algorithmic}
\end{algorithm}
\vspace{-1em}
Each block is then classified into one of four cases:
\begin{itemize}[leftmargin=3.5mm]
    \item \textbf{Divergence (line~\ref{line:divergence_branch}):} If the residual exceeds the ceiling $\mathrm{Res}_{\text{ceil}} = 10^{2}$, the block is considered to be diverging.
    This casts an immediate REMOVE vote ($\Delta_{\text{local}} = -P_{\text{node}}$), bypassing \textsc{Hysteresis} entirely and resetting both hysteresis
    counters, since a diverging block should not wait for consecutive votes.
    \item \textbf{Fast convergence (line~\ref{line:fast_branch}):} If the block converged with both $\rho_{avg} \leq \rho_{\text{fast}}$ and $\eta_k \leq \alpha_{\text{fast}}$, 
    it casts a fast-convergence vote through \textsc{Hysteresis}. This becomes an ADD vote ($\Delta_{\text{local}} = +P_{\text{node}}$) once $H_{\text{fast}} = 2$ consecutive 
    blocks register fast convergence, thereby triggering a REPLACE PSetOp with a $P_{add}$ PSet. While the count is still accumulating, the vote remains neutral ($\Delta_{\text{local}} = 0$).
    \item \textbf{No convergence (line~\ref{line:slow_branch}):} If the block exhausts its iteration budget ($k = K^{\max}$) without converging and $\rho_{avg} \geq \rho_{\text{slow}}$,
     it casts a slow-convergence vote through \textsc{Hysteresis}. This becomes a REMOVE vote ($\Delta_{\text{local}} = -P_{\text{node}}$) once $H_{\text{slow}} = 2$ consecutive blocks 
     register slow convergence, triggering a REPLACE PSetOp with a $P_{sub}$ PSet. Otherwise, it casts a $\text{slowwait}$ vote ($\Delta_{\text{local}} = 0$, $\text{slowwait} = \text{true}$), 
     indicating that the block is genuinely struggling before a REMOVE vote is triggered.
    \item \textbf{Slow convergence (line~\ref{line:default_branch}):} In all other cases, such as when the block converges without meeting the fast criteria, 
    or has exhausted its iteration budget, it casts a neutral KEEP vote ($\Delta_{\text{local}} = 0$, $\text{slowwait} = \text{false}$), and the block size remains unchanged. 
    The hysteresis counters are reset once the block finishes (either by converging or by reaching $K^{\max}$), ensuring that the next vote starts fresh.
\end{itemize}
\vspace{-0.75em}
\begin{algorithm}
\caption{\textsc{parallel\_steps\_resizer}($N_{ts}, K^{\max}, \mathcal{H}_b$)}
\label{alg:parallel_steps_resizer}
\begin{algorithmic}[1]
    \State $\rho_{avg} \gets \left(\mathrm{Res}^{k} / \mathrm{Res}^{1}\right)^{1/k}$ \label{line:rho_avg}
    \State $\eta_k \gets k / K^{\max}$ \label{line:iterratio}
    \State $\text{converged} \gets \big(\mathrm{Res}^k < \varepsilon_{\text{tol}}\big)$ \label{line:converged}
    \State $\Delta \gets 0$
    \State $\text{slowwait} \gets \text{false}$

    \hspace{\algorithmicindent} \CommentAbove{\textit{Divergence}}
    \If{($\mathrm{Res}^{k} > \mathrm{Res}_{\text{ceil}}$)} \label{line:divergence_branch}
        \State $\Delta \gets -P_{\text{node}}$
        \State $c_{\text{fast}}, c_{\text{slow}} \gets 0, 0$
    \hspace{\algorithmicindent} \CommentAbove{\textit{High rate of convergence}}
    \ElsIf{($\text{converged}$ \textbf{\&\&} $\eta_k \leq \alpha_{\text{fast}}$ \textbf{\&\&} $\rho_{avg} \leq \rho_{\text{fast}})$} \label{line:fast_branch}
        \State $\Delta \gets \Call{Hysteresis}{+P_{\text{node}},\ c_{\text{fast}},\ H_{\text{fast}}}$
        \State $c_{\text{slow}} \gets 0$
    \hspace{\algorithmicindent} \CommentAbove{\textit{No convergence}}
    \ElsIf{($k == K^{\max}$ \textbf{\&\& not} $\text{converged}$ \textbf{\&\&} $\rho_{avg} \geq \rho_{\text{slow}}$)} \label{line:slow_branch}
        \State $\Delta \gets \Call{Hysteresis}{-P_{\text{node}},\ c_{\text{slow}},\ H_{\text{slow}}}$
        \State $c_{\text{fast}} \gets 0$
        \State $\text{slowwait} \gets \text{true}$
    \hspace{\algorithmicindent} \CommentAbove{\textit{Slow convergence}}
    \ElsIf{($\text{converged}$ \textbf{||} $k == K^{\max}$)} \label{line:default_branch}
            \State $c_{\text{fast}}, c_{\text{slow}} \gets 0, 0$
    \EndIf

    \State $\Delta \gets \Call{ReconcileResizeDelta}{\Delta,\ \text{slowwait}}$
    \State \Return $\Delta$
\end{algorithmic}
\end{algorithm}

Each rank, namely, the MPI process running one of the $N_{ts}$ parallel timesteps in the block (i.e.\ a processor $P_n$ in Alg.~\ref{alg:pfasst_iteration}), derives
$\Delta$ from its own local $\mathcal{H}_b$. As previously established, correction information propagates sequentially across
timesteps, so later timesteps in a block may need more iterations than earlier ones, meaning that different ranks can vote differently in a block.
\begin{algorithm}[H]
\caption{\textsc{ReconcileResizeDelta}($\Delta_{\text{local}}$, $\text{slowwait}_{\text{local}}$)}
\label{alg:reconcile}
\begin{algorithmic}[1]
\State $\text{anyRemove} \gets \Call{AllReduceMax}{\Delta_{\text{local}} < 0}$
\State $\text{anySlowWait} \gets \Call{AllReduceMax}{\text{slowwait}_{\text{local}}}$
\If{$\text{anyRemove}$}
    \State \Return $-P_{\text{node}}$ 
\ElsIf{$\text{anySlowWait}$}
    \State \Return $\min(\Delta_{\text{local}}, 0)$ 
\Else
    \State \Return \Call{Broadcast}{$\Delta_{\text{local}}$ of 0} 
\EndIf
\end{algorithmic}
\end{algorithm}
The \textsc{parallel\_steps\_resizer} routine therefore reconciles these local votes into one global $\Delta$ using
\textsc{ReconcileResizeDelta} given in Alg.~\ref{alg:reconcile}. A REMOVE vote from any single rank overrides any ADD or KEEP vote. 
A genuine slowwait from any rank suppresses all ADD votes. Otherwise, when no rank votes REMOVE or reports a slowwait, rank~$0$'s
vote decides whether the block grows or stays the same.
\subsection{Iteration-Ceiling Adaptation}
\textsc{parallel\_steps\_resizer} has a fixed iteration ceiling $K^{\max}$ and only the block size $N_{ts}$ is adapted.
\textsc{iters\_resizer} extends \textsc{parallel\_steps\_resizer} by using the reconciled $\Delta$
to re-anchor the iteration ceiling to the new block size.
As established above, Parareal-type schemes are known to converge exactly within $N_{ts}$ iterations for linear 
problems~\cite{50Years}; however, PFASST does not share this guarantee~\cite{unifiedpint}, so its required number of 
iterations cannot be bounded a priori. In previous work, the iteration ceiling for PFASST was also determined empirically by 
incrementally increasing the iteration count until the scheme achieved sufficient accuracy and stability~\cite{pfasstsh}.
We therefore set the maximum iteration ceiling, $K^{\max}_{\text{ceil}}$, equal to $N_{ts}^{\max}$, which is the largest block
size possible across our experiments, i.e., the maximum value to which the resizer is allowed to grow, rather than the largest block size encountered in any single run. 
This approach ensures that the resizer has enough budget to enable convergence even in the worst case.
\vspace{-1em}
\begin{algorithm}[H]
\caption{\textsc{iters\_resizer}($N_{ts}, K^{\max}, \mathcal{H}_b$)}
\label{alg:iters_resizer}
\begin{algorithmic}[1]
    \State $\Delta \gets \Call{parallel\_steps\_resizer}{N_{ts}, K^{\max}, \mathcal{H}_b}$
    \State $\tilde{N}_{ts} \gets N_{ts} + \Delta$
    \State $\tilde{K}^{\max}_{\text{local}} \gets \min\big(\max(K^{\max}_{\text{floor}},\, \tilde{N}_{ts}),\ K^{\max}_{\text{ceil}}\big)$
    \State $\tilde{K}^{\max} \gets \Call{AllReduceMin}{\tilde{K}^{\max}_{\text{local}}}$
    \State \Return $(\Delta,\ \tilde{K}^{\max})$
\end{algorithmic}
\end{algorithm}
\vspace{-1.5em}
We also define a minimum iteration ceiling, $K^{\max}_{\text{floor}} = 10$, 
to serve as a floor throughout this work. The new local iteration ceiling $\tilde{K}^{\max}_{\text{local}}$ is then the larger of the floor
$K^{\max}_{\text{floor}}$ and the target parallelism $\tilde{N}_{ts}$, clamped from above at $K^{\max}_{\text{ceil}}$. Since different ranks can reach this decision at different real times, each rank's locally computed target ceiling may differ,
even when $\Delta$ is already unanimous. A mismatched $K^{\max}$ across ranks is not merely a lost-accuracy risk but a deadlock risk. $K^{\max}$
bounds how many rounds of status exchange each rank expects from its neighbors, so a mismatch can leave one rank's send/receive chain out of
sync with its neighbor's. To prevent this, \textsc{iters\_resizer} performs a global reduction to take the minimum ceiling that any rank independently
derived, ensuring agreement. This reconciled $K^{\max}$ is then returned alongside $\Delta$.

\section{Numerical Experiments}
We evaluate the adaptive PFASST framework introduced in 
Section~\ref{sec:adaptiveframework} using
the \textsf{SWEET} library~\cite{sweet}, which provides implementations of
well-established numerical test cases for the global Shallow Water Equations (SWE) in spherical
geometry. We compare the static (i.e., non-adaptive) PFASST configuration~\cite{pfasstsh} against
\textsc{parallel\_steps\_resizer} and \textsc{iters\_resizer} to assess whether convergence-driven
resource adaptation improves resource usage, wall-clock time, and the overall cost-performance
tradeoff. Section~\ref{sec:swe} describes the governing equations, while
Section~\ref{sec:results} presents results for two benchmarks: the advection of a cosine bell along
the equator and the nonlinear evolution of an unstable barotropic wave.
\subsection{Shallow Water Equations on the Sphere}\label{sec:swe}
The rotating SWE provide a simplified yet representative model for the horizontal dynamics of atmospheric flows, 
and have been used to investigate parallel-in-time integration schemes for numerical weather prediction and for 
barotropic solvers in ocean models~\cite{sdcsphere, mlsdcsh, pfasstsh}. To avoid
the coordinate singularity that arises at the poles when the equations are formulated in terms of
velocity components, we adopt the vorticity--divergence formulation, with prognostic variables
$U = (\phi, \zeta, \delta)$ denoting the geopotential, relative vorticity, and horizontal divergence,
respectively. The governing equations~\cite{pfasstsh} take the form
\begin{align} \label{eq:swe_split}
    \frac{\partial U}{\partial t} = \begin{bmatrix} -\bar{\phi}\delta + \nu\nabla^2\phi' \\ 
        \nu\nabla^2\zeta \\ 
        -\nabla^2\phi + \nu\nabla^2\delta \\
        \end{bmatrix} + \begin{bmatrix} -\nabla\cdot(\phi'V)\\
         -\nabla\cdot(\zeta+f)V \\
        k\cdot\nabla \times (\zeta+f)V - \nabla^2(\frac{V\cdot V}{2})\\
        \end{bmatrix}
\end{align}
where the first vector of the right-hand side sum is a collection of stiff, linear terms associated with fast gravity-wave propagation and
diffusion, and the second vector contains non-stiff, nonlinear terms associated with advection and the
Coriolis force. 
This split motivates an implicit--explicit (IMEX) treatment: linear terms are integrated
implicitly, since resolving gravity waves explicitly would impose a severe CFL restriction on the
timestep, while nonlinear terms are integrated explicitly, avoiding the cost of a nonlinear implicit solve.
We discretize \eqref{eq:swe_split} in space using a Spherical Harmonics (SH) spectral
transform and employ PFASST for time integration~\cite{sweet}. For clarity, Alg.~\ref{alg:pfasst_iteration}
presents the SDC sweep in its fully implicit form; in practice, we use the IMEX
extension~\cite{minion2003} to treat the linear and nonlinear terms of \eqref{eq:swe_split}
implicitly and explicitly, respectively.

\subsection{Results}\label{sec:results}
All experiments are conducted on the SuperMUC-NG~\cite{supermuc}, using \dynres{} to
reallocate compute nodes to a running PFASST block at runtime. SuperMUC-NG consists of 6{,}336 thin
compute nodes, each equipped with Intel Skylake Xeon Platinum 8174 processors with 48 cores and
96 GB of memory per node. 
For both benchmarks, we exploit spatial parallelism via OpenMP with
\texttt{OMP\_NUM\_THREADS}=48, and temporal parallelism via
MPI, allocating each parallel timestep to a single thin compute node.
Application-specific parameters used for both benchmarks are given in Table~\ref{tab:common_params}.
As established in Section~\ref{sec:adaptiveframework}, we use 2 PFASST levels with 1 SDC sweep per
level. 
\begin{table}[h]
    \centering
    \footnotesize
    \vspace{-0.5em}
    \begin{tabular}{||l|l||}
        \hline
        Parameter & Value \\
        \hline
        Spatial resolution $N$ & 256 SH modes\\
        Timestep $\Delta t$ & 240\,s \\
        SDC node type & Gauss--Lobatto \\
        SDC nodes [$M_f+1, M_c+1$] & [5, 3] \\
        Spatial coarsening ratio $\alpha$ & 0.6 \\
        \hline
    \end{tabular}
    \vspace{0.5em}
    \caption{Application-specific parameters}
    \label{tab:common_params}
    \vspace{-2.5em}
\end{table}
Our choice of spatial coarsening ratio $\alpha=0.6$, together with $M_f+1=5$ and $M_c+1=3$
SDC nodes, matches the configuration identified as achieving stable timestep size of
up to 240s while still providing computational savings~\cite{pfasstsh}.
For each benchmark, we compare the static PFASST baseline, fixed
at a block size of $N_{ts}=32$, against both \textsc{parallel\_steps\_resizer} and
\textsc{iters\_resizer}, each of which starts from three initial block sizes $N_{ts} \in \{1, 16, 32\}$.
We limit the largest possible block size to 32, and similarly cap the iteration ceiling at
$K^{\max}=32$, because larger block sizes tend to destabilize PFASST. Additionally, we set
$P_{\text{node}}=1$. Since each timestep is assigned to a single compute node, increasing or decreasing
the block by $P_{\text{node}}$ corresponds to adding or removing one compute node at a time.

\subsubsection*{Zonal Advection of a Cosine Bell}\label{sec:williamson1}
\begin{figure}[h]
    \centering
    \includegraphics[width=\columnwidth]{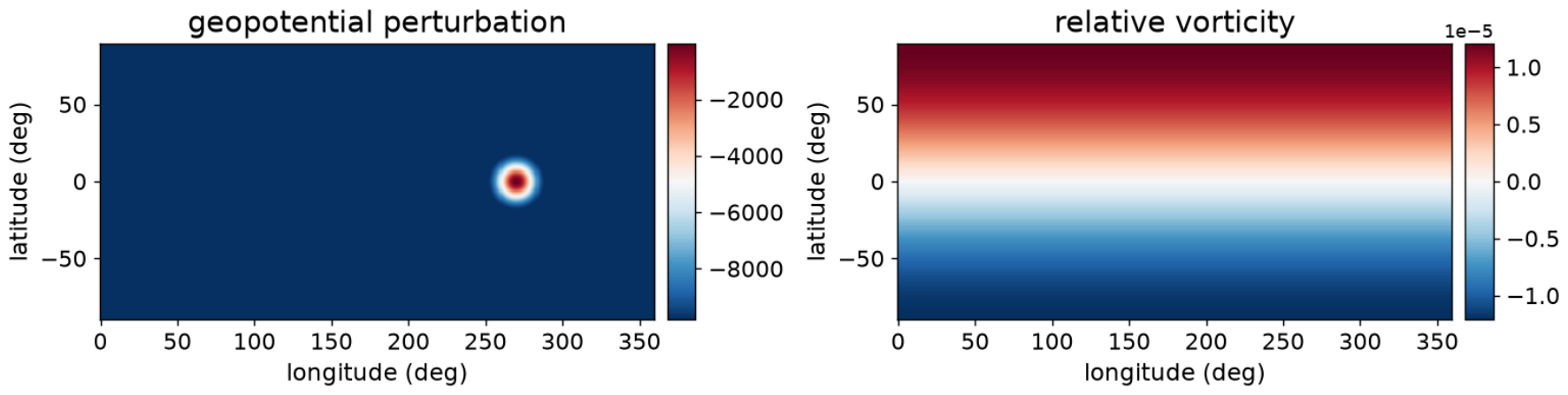}
    \vspace{-2em}
    \caption{Geopotential perturbation $\phi'$ (left) and relative vorticity $\zeta$ (right) at the end
    of the 12-day simulation.}
    \label{fig:williamson1_solution}
\end{figure}

\begin{figure}[t]
    \centering
    \includegraphics[width=\columnwidth]{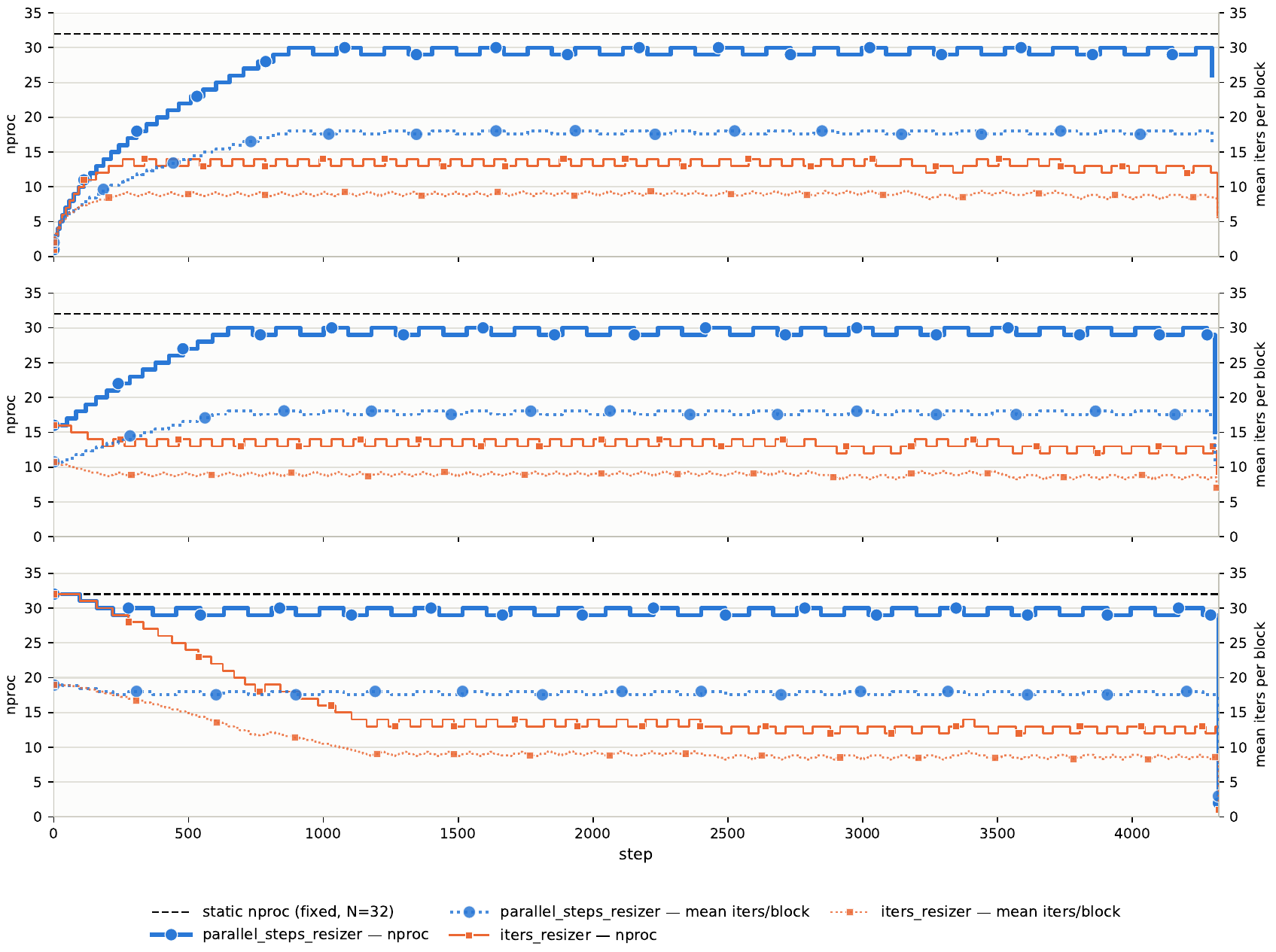}
    \vspace{-1.5em}
    \caption{Number of allocated processes (nproc) and mean SDC iterations per block over the course
    of the simulation for \textsc{parallel\_steps\_resizer} and \textsc{iters\_resizer}, started from
    initial block sizes $N_{ts}\in\{1,16,32\}$ (top to bottom), compared against the static baseline
    (dashed line).}
    \label{fig:williamson1_resource_adaptation}
    \vspace{-1em}
\end{figure}

\begin{figure*}[h]
    \centering
    \includegraphics[width=0.8\textwidth]{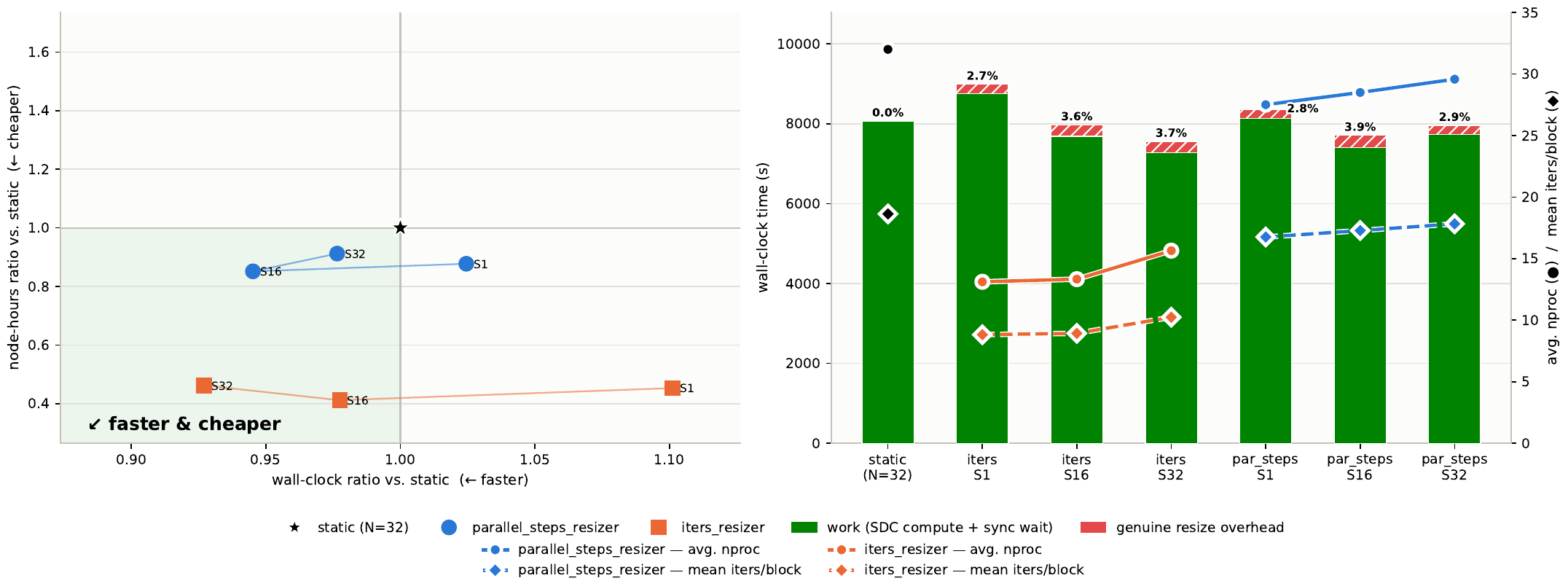}
    \vspace{-1em}
    \caption{Left: wall-clock time and node-hours (note the differing scales of the axes), both normalized to the static baseline, for
    \textsc{parallel\_steps\_resizer} and \textsc{iters\_resizer} started from $N_{ts}\in\{1,16,32\}$ (labeled S1, S16, S32);
    the shaded region marks configurations that are both faster and cheaper than static case.
    Right: wall-clock time broken down into SDC work and genuine resize overhead (stacked bars, with
    the overhead share annotated above each bar), alongside
    the time-averaged number of allocated processes and mean SDC iterations per block (lines, right
    axis) for the same configurations.}
    \label{fig:williamson1_summary}
    \vspace*{-1em}
\end{figure*}

The first benchmark advects a cosine bell once around the sphere on a prescribed, solid-body zonal
wind field, following the initial condition proposed by Williamson et al.~\cite{williamson1}. Since
the wind field is derived entirely from a stream function, it is non-divergent by construction
($\delta\equiv0$) and prescribed independently of the height field, so the exact solution is simply
the initial bell translated along the equator without distortion as shown in Fig.~\ref{fig:williamson1_solution}. 
We run the benchmark for 12 days,
the time needed for the bell to complete exactly one revolution, which allows this benchmark to
isolate errors in the nonlinear, advective part of the discretization from those associated with 
stiff gravity-wave dynamics.
\vspace{-0.5em}
\vspace{-0.75em}
Fig.~\ref{fig:williamson1_resource_adaptation} shows the evolution of block size over the simulation for both resizing
strategies along with the gray dashed line representing an unchanging block size of 32 for the static case. 
The \textsc{parallel\_steps\_resizer} strategy grows up to $N_{ts}$=30 and then oscillates around a block size of 29 
and 30 until the end of the simulation.
At $N_{ts}=30$, the last rank, which waits longest in PFASST's serial coarse-sweep pipeline, is structurally slowest to converge and,
therefore, is typically the first whose $\rho_{avg}$ reaches $\rho_{\text{slow}}$. 
Since a single REMOVE vote overrides the rest, this one slow rank forces a shrink back to $N_{ts}=29$. 
Therefore, blocks oscillate near their marginally fast block size of 30. 
\textsc{iters\_resizer} also stabilizes for similar reasons but at a lower block size of around 13 or 14  
since $K^{\max}_{\text{ceil}}$ is also adjusted in conjunction with the block size.
The corresponding average iterations per block for each strategy also oscillate but much lower than the block size at \textasciitilde16 and 
\textasciitilde9 for \textsc{parallel\_steps\_resizer} and \textsc{iters\_resizer} respectively.
The mean masks a notable spread, however: the per-block maximum iteration count can be significantly higher than the mean, and
can even exceed the block size itself, since a single slow-converging rank may still demand close to the full iteration
ceiling regardless of how small the current block is.
In Fig.~\ref{fig:williamson1_summary}, we compute wall-clock times and node-hour ratios with respect to 
the static case with a block size of 32. 
For both resizers, starting from $N_{ts}=16$ provides the best cost-performance tradeoff.
It is both faster and cheaper than the static baseline, while starting from $N_{ts}=1$ incurs a wall-clock penalty as the block size incrementally
increases from a single timestep. \textsc{iters\_resizer} delivers significantly greater cost savings than \textsc{parallel\_steps\_resizer}, achieving
up to 59\% fewer node-hours compared to up to 15\% while maintaining a similar wall-clock cost, since re-anchoring $K^{\max}_{\text{ceil}}$ to the
shrinking block size lets it settle at a much lower equilibrium than \textsc{parallel\_steps\_resizer}'s fixed ceiling allows.
In its best case ($N_{ts}=16$), \textsc{iters\_resizer} uses an average block size of 13.3 and a mean of 8.9 iterations per block, both less than
half of what the static case consumes (32 timesteps, 18.6 mean iterations).
\begin{table}[h]
    \centering
    \footnotesize
    \begin{tabular}{||l|l|l|l||}
        \hline
        Strategy & $N_{ts}$ & Advection & Barotropic wave \\
        \hline
        static & 32 & $6.45\times10^{-2}$ & $9.94\times10^{-5}$ \\
        \textsc{parallel\_steps\_resizer} & 1 & $6.58\times10^{-4}$ & $2.31\times10^{-4}$ \\
        \textsc{parallel\_steps\_resizer} & 16 & $7.49\times10^{-4}$ & $1.45\times10^{-4}$ \\
        \textsc{parallel\_steps\_resizer} & 32 & $3.76\times10^{-3}$ & $1.16\times10^{-4}$ \\
        \textsc{iters\_resizer} & 1 & $1.57\times10^{-1}$ & $1.16\times10^{-4}$ \\
        \textsc{iters\_resizer} & 16 & $1.58\times10^{-1}$ & $1.11\times10^{-4}$ \\
        \textsc{iters\_resizer} & 32 & $1.63\times10^{-1}$ & $1.00\times10^{-4}$ \\
        \hline
    \end{tabular}
    \vspace{0.5em}
    \caption{RMS error of the final-timestep solution against the static, serial ($N_{ts}=1$) reference, for both benchmarks}
    \label{tab:accuracy}
    \vspace{-3em}
\end{table}
We report the accuracy of each configuration in Table~\ref{tab:accuracy} by comparing the final-timestep solution
against the static, serial ($N_{ts}=1$) reference.
\textsc{parallel\_steps\_resizer} is substantially more
accurate than the static $N_{ts}=32$ baseline, consistent with PFASST's accuracy degrading as more timesteps are computed in
parallel. Since it runs with a slightly smaller average block size than the fixed $N_{ts}=32$ static case, it accumulates less
of this degradation. \textsc{iters\_resizer}, in contrast, is less accurate than the static run despite an even smaller average block
size. Its aggressively reduced iteration budget evidently costs more accuracy than the smaller block size recovers. The two
resizers therefore occupy different points on a genuine cost-accuracy trade-off: \textsc{parallel\_steps\_resizer} improves on the
static run's accuracy alongside modest cost savings, while \textsc{iters\_resizer} trades a larger accuracy loss for substantially
larger resource savings. Additionally, the actual overhead associated with resizing
remains below 4\% of the total wall-clock time for every configuration. This is because the underlying PSetOp is issued as a non-blocking MPI request and
resize checks only poll its completion via \texttt{MPI\_Test}. Therefore, the only cost incurred is the one-time communicator rebuild and state synchronization once a change
is actually applied, consistent with the reconfiguration-overhead benchmarks reported for this DPP mechanism~\cite{drmheat}.
\begin{figure*}[h]
    \centering
    \includegraphics[width=0.8\textwidth]{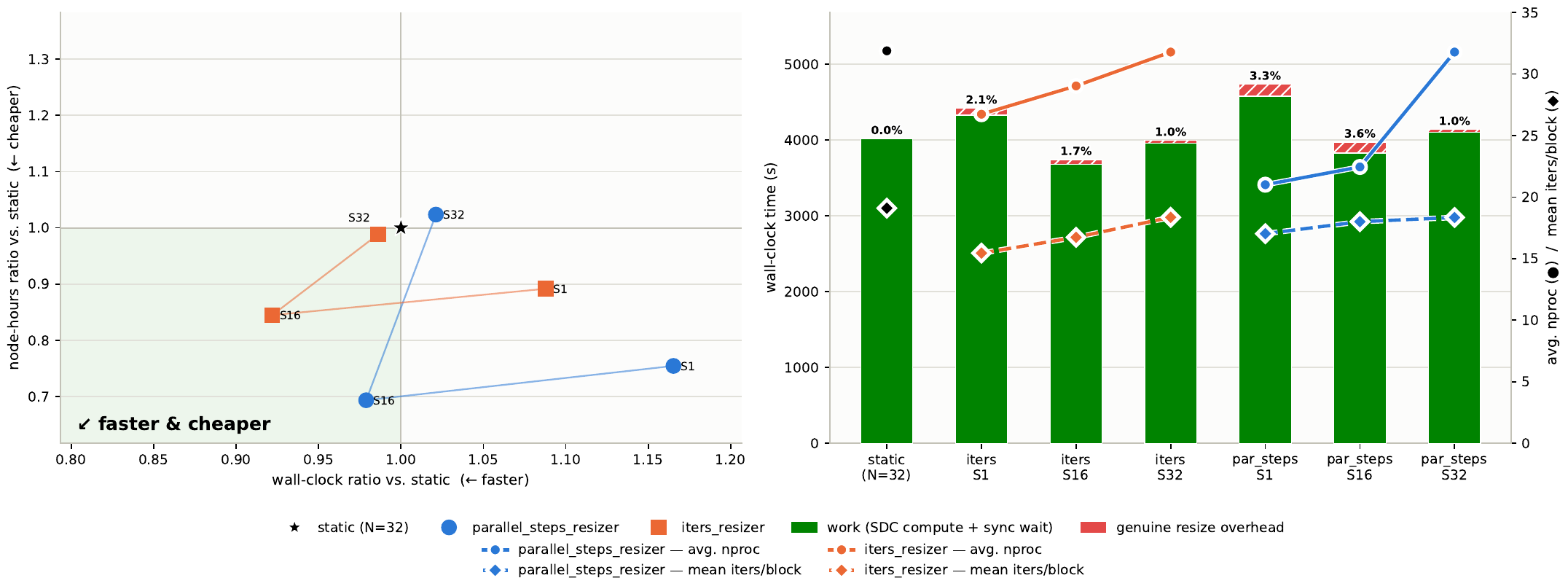}
    \vspace{-1em}
    \caption{Left: wall-clock time and node-hours (note the differing scales of the axes), both normalized to the static baseline, for
    \textsc{parallel\_steps\_resizer} and \textsc{iters\_resizer} started from $N_{ts}\in\{1,16,32\}$ (labeled S1, S16, S32);
    the shaded region marks configurations that are both faster and cheaper than static.
    Right: wall-clock time broken down into SDC work and genuine resize overhead (stacked bars, with
    the overhead share annotated above each bar), alongside
    the time-averaged number of allocated processes and mean SDC iterations per block (lines, right
    axis) for the same configurations.}
    \label{fig:galewsky_summary}
    \vspace{-1em}
\end{figure*}

\subsubsection*{Unstable Barotropic Wave}\label{sec:galewsky}
The second benchmark is based on the unstable zonal jet perturbed by a localized Gaussian bump in the geopotential field~\cite{galewsky}. The unperturbed jet is a geostrophically balanced, 
steady-state solution of the SWE. The added perturbation initiates a barotropic instability, 
causing the jet to roll up into vortices over the course of the 6-day simulation. 
We set the diffusion coefficient to $\nu=10^5\,\text{m}^2\text{s}^{-1}$
to stabilize the flow dynamics and reduce the errors caused by
under-resolved nonlinear interaction modes, as a cheaper alternative to spectral viscosity~\cite{pfasstsh}.
This benchmark exhibits a markedly different resource adaptation trend than the advection benchmark. 
\begin{figure}[H]
    \centering
    \includegraphics[width=\columnwidth]{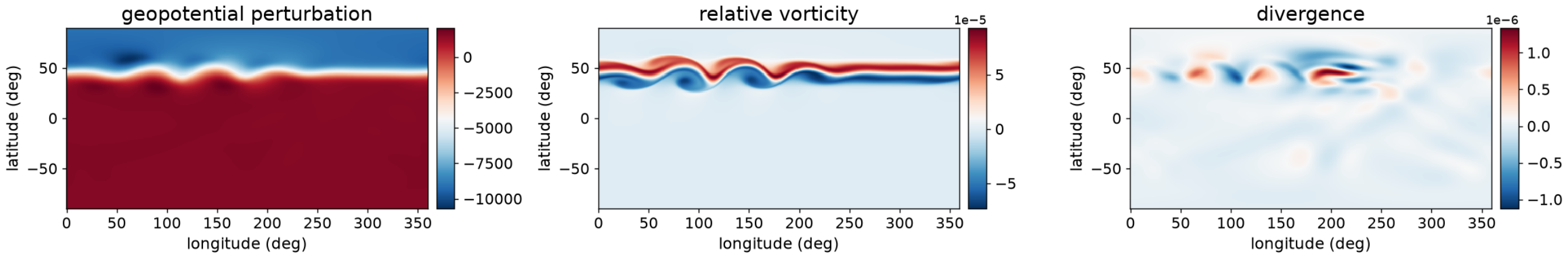}
    \vspace{-2em}
    \caption{Geopotential perturbation $\phi'$, relative vorticity $\zeta$, and divergence $\delta$
    (left to right) after 6 days of simulation.}
    \label{fig:galewsky_solution}
\end{figure}
\vspace{-1.25em}
\begin{figure}[H]
    \centering
    \includegraphics[width=\columnwidth]{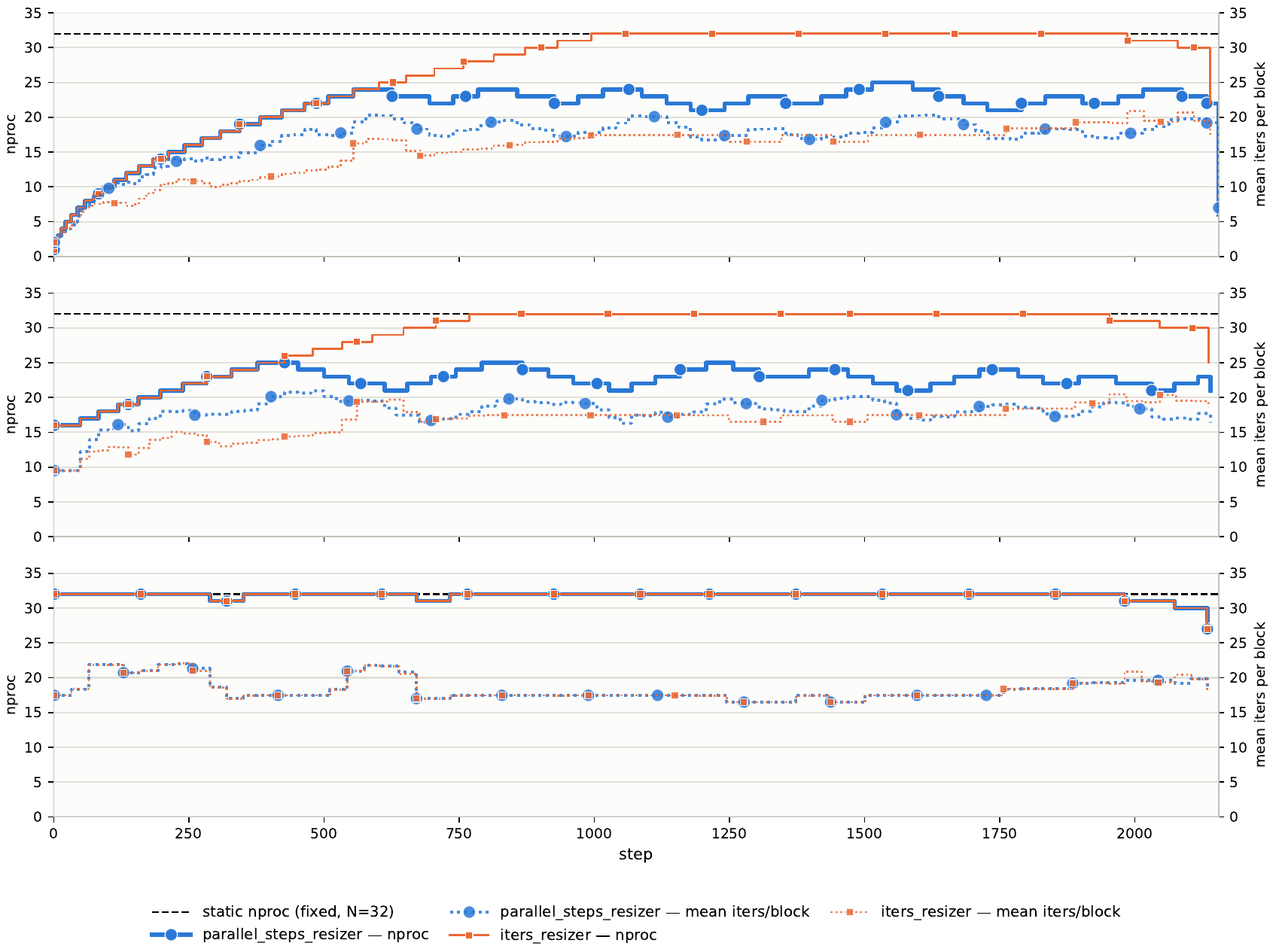}
    \vspace{-1.5em}
    \caption{Number of allocated processes (nproc) and mean SDC iterations per block over the course
    of the simulation for \textsc{parallel\_steps\_resizer} and \textsc{iters\_resizer}, started from
    initial block sizes $N_{ts}\in\{1,16,32\}$ (top to bottom), compared against the static baseline
    (dashed line).}
    \label{fig:galewsky_resource_adaptation}
\end{figure}
\vspace{-1em}
As shown in Fig.~\ref{fig:galewsky_resource_adaptation}, \textsc{iters\_resizer} grows all the way to the maximum 
allowed block size ($N_{ts}=32$) for all starting block sizes, while \textsc{parallel\_steps\_resizer} grows only to around 
$N_{ts}=24$ for the case starting from 1 timestep and $N_{ts}=25$ for the case starting from 16 timesteps, before oscillating 
for the remainder of the simulation.
This difference arises from the benchmark's nonlinear dynamics. The block frequently 
requires more iterations than its current size can support from early in the simulation, causing the residual to grow (not shown here) with \textsc{parallel\_steps\_resizer}. 
This growth triggers repeated REMOVE votes, reducing the block size until fast convergence is restored, only for the block to 
subsequently grow back into the same mismatch, resulting in an oscillatory pattern that persists until the end of the simulation. 
When starting from $N_{ts}=32$, the configuration consistently uses the full 32-iteration budget from the outset and remains 
perpetually on the brink of instability. We also test this configuration with a relaxed iteration ceiling of 
$K^{\max}_{\text{ceil}}=48$ (not shown), but the block is only able to grow marginally before becoming unstable and shrinking again. 
\textsc{iters\_resizer} exhibits a similar pattern, but the instability and resulting REMOVE operations occur primarily near the 
end of the simulation, when some blocks become unstable. Although the adaptations differ from the advection benchmark, starting with $N_{ts}=16$ gives the best trade-off for both resizers.
Starting from $N_{ts}=1$ incurs an even larger wall-clock penalty than for the advection benchmark, as the block ramps up from a
single timestep: up to 16.5\% slower for \textsc{parallel\_steps\_resizer} and 8.8\% slower for \textsc{iters\_resizer} than static.
While $N_{ts}=32$ already starts at the configured ceiling, for \textsc{parallel\_steps\_resizer} this configuration is actually
2.1\% slower and 2.4\% more expensive than static, rather than merely tracking it. \textsc{parallel\_steps\_resizer} reduces node-hours by up to 31\% and \textsc{iters\_resizer} by up to 16\% at $N_{ts}=16$,
both while also reducing wall-clock time there and, as with the advection benchmark, this is attributed to the reduced number of
iterations and resources used. At $N_{ts}=16$ (the best case for wall-clock time), \textsc{iters\_resizer} averages 29.0 processes 
and 16.7 iterations per block, whereas \textsc{parallel\_steps\_resizer} averages 22.4 processes and 18.0 iterations, 
compared to 31.9 processes and 19.1 mean iterations for the static baseline.
We evaluate the accuracy of each configuration using the same static, serial ($N_{ts}=1$) reference as for the advection
benchmark in Table~\ref{tab:accuracy}. Unlike the advection benchmark,
no resizer stands out here: the RMS error against this reference is of order $10^{-4}$ across all configurations,
including the static $N_{ts}=32$ baseline, indicating that resource adaptation does not introduce error beyond the
parallel-in-time error already inherent to running multiple timesteps concurrently.
Resize overhead again stays below 4\% of total wall-clock time across all
configurations, as can be seen in Fig.~\ref{fig:galewsky_summary}.

\section{Conclusion}
Iterative parallel-in-time algorithms require selecting optimal parameters, which are often problem-dependent and typically unknown prior to running a simulation. 
In particular, determining the optimal number of parallel-in-time steps frequently forces users to rely on costly trial-and-error tuning.

In this work, we use the PFASST parallel-in-time method to demonstrate how dynamically adjusting the computational resources at runtime can address this challenge.
This is achieved by using \dynres~software to extend PFASST with support for varying computing resources at runtime.
We develop and investigate two convergence-driven strategies that dynamically adapt the parallel timesteps and, consequently, the number of compute nodes, with one strategy additionally adjusting the iteration ceiling.
Resource changes, including the addition of new processes, are applied asynchronously, allowing newly added processes to initialize their own state without requiring data redistribution while existing processes continue their parallel-in-time computations uninterrupted.

The block size trends observed for each strategy differed significantly between two PDE benchmarks, 
reflecting genuinely distinct convergence characteristics. This highlights the need for PFASST to employ strategies that allow it to tune itself at runtime based on its own convergence behavior, 
rather than relying on a fixed static configuration.
Despite these differing trends, both strategies incurred negligible resizing overhead due to the asynchronous implementation, remaining below 4\% of total wall-clock time for both benchmarks. Both approaches also consistently 
require fewer average processes and iterations than the corresponding static runs, reducing node-hours by up to 59\%
for the advection benchmark and up to 31\% for the barotropic wave benchmark, all while matching the wall-clock time 
and the accuracy of the static baseline.



\bibliographystyle{ACM-Reference-Format}
\bibliography{bibliography}

@article{minion2003,
  author  = {Minion, Michael L.},
  title   = {Semi-implicit spectral deferred correction methods for ordinary differential equations},
  journal = {Communications in Mathematical Sciences},
  volume  = {1},
  number  = {3},
  pages   = {471--500},
  year    = {2003}
}

@article{Alya,
  author  = {V{\'a}zquez, M. and Houzeaux, G. and Koric, S. and Artigues, A. and Aguado-Sierra, J. and Ar{\'i}s, R. and Mira, D. and Calmet, H. and Cucchietti, F. and Owen, H. and Taha, A. and Burness, E. D. and Cela, J. M. and Valero, M.},
  title   = {Alya: Multiphysics engineering simulation towards exascale},
  journal = {Journal of Computational Science},
  volume  = {14},
  pages   = {15--27},
  year    = {2016},
  doi     = {10.1016/j.jocs.2015.12.007}
}

@article{gromacs,
  author  = {Abraham, Mark James and Murtola, Teemu and Schulz, Roland and P{\'a}ll, Szil{\'a}rd and Smith, Jeremy C. and Hess, Berk and Lindahl, Erik},
  title   = {{GROMACS}: High performance molecular simulations through multi-level parallelism from laptops to supercomputers},
  journal = {SoftwareX},
  volume  = {1-2},
  pages   = {19--25},
  year    = {2015},
  doi     = {10.1016/j.softx.2015.06.001}
}

@article{houzeaux2022,
  author  = {Houzeaux, Guillaume and Badia, Rosa M. and Borrell, Ricard and Dosimont, Damien and Ejarque, Jorge and Garcia-Gasulla, Marta and L{\'o}pez, Vicen{\c{c}}},
  title   = {Dynamic resource allocation for efficient parallel {CFD} simulations},
  journal = {Computers \& Fluids},
  volume  = {245},
  pages   = {105577},
  year    = {2022},
  doi     = {10.1016/j.compfluid.2022.105577}
}

@misc{sandas2026,
  author = {Sand{\aa}s, Petter and Iserte, Sergio and Ar{\'e}jula-A{\'i}sa, {\'I}{\~n}igo and Hess, Berk and Pe{\~n}a, Antonio J.},
  title  = {Malleable Molecular Dynamics Simulations with {GROMACS} and {DMR}},
  year   = {2026},
  eprint = {2605.14655},
  archivePrefix = {arXiv}
}

@article{sdc,
  author  = {Dutt, Alok and Greengard, Leslie and Rokhlin, Vladimir},
  title   = {Spectral Deferred Correction Methods for Ordinary Differential Equations},
  journal = {BIT},
  year    = {2000},
  month   = jun,
  volume  = {40},
  pages   = {241--266},
  doi     = {10.1023/A:1022338906936}
}

@article{mlsdc,
  author    = {Speck, Robert and Ruprecht, Daniel and Emmett, Matthew and Minion, Michael and Bolten, Matthias and Krause, Rolf},
  title     = {A Multi-Level Spectral Deferred Correction Method},
  journal   = {BIT Numerical Mathematics},
  year      = {2014},
  month     = aug,
  volume    = {55},
  number    = {3},
  pages     = {843--867},
  issn      = {1572-9125},
  publisher = {Springer Science and Business Media LLC},
  doi       = {10.1007/s10543-014-0517-x},
  url       = {http://dx.doi.org/10.1007/s10543-014-0517-x}
}

@article{mlsdcsh,
  author   = {Hamon, Fran{\c{c}}ois P. and Schreiber, Martin and Minion, Michael L.},
  title    = {Multi-Level Spectral Deferred Corrections Scheme for the Shallow Water Equations on the Rotating Sphere},
  journal  = {Journal of Computational Physics},
  year     = {2019},
  volume   = {376},
  pages    = {435--454},
  issn     = {0021-9991},
  doi      = {10.1016/j.jcp.2018.09.042},
  url      = {https://www.sciencedirect.com/science/article/pii/S0021999118306442}
}

@article{sdcsphere,
  author    = {Jia, Jun and Hill, Judith C. and Evans, Katherine J. and Fann, George I. and Taylor, Mark A.},
  title     = {A Spectral Deferred Correction Method Applied to the Shallow Water Equations on a Sphere},
  journal   = {Monthly Weather Review},
  year      = {2013},
  publisher = {American Meteorological Society},
  address   = {Boston, MA, USA},
  volume    = {141},
  number    = {10},
  pages     = {3435--3449},
  doi       = {10.1175/MWR-D-12-00048.1},
  url       = {https://journals.ametsoc.org/view/journals/mwre/141/10/mwr-d-12-00048.1.xml}
}

@inproceedings{pararealswe,
  author    = {Arbenz, Peter and Hiltebrand, Andreas and Obrist, Dominik},
  editor    = {Wyrzykowski, Roman and Dongarra, Jack and Karczewski, Konrad and Wa{\'s}niewski, Jerzy},
  title     = {A Parallel Space-Time Finite Difference Solver for Periodic Solutions of the Shallow-Water Equation},
  booktitle = {Parallel Processing and Applied Mathematics},
  year      = {2012},
  publisher = {Springer Berlin Heidelberg},
  address   = {Berlin, Heidelberg},
  pages     = {302--312},
  isbn      = {978-3-642-31500-8}
}

@article{pfasstsh,
  author    = {Hamon, Fran{\c{c}}ois P. and Schreiber, Martin and Minion, Michael L.},
  title     = {Parallel-in-Time Multi-Level Integration of the Shallow-Water Equations on the Rotating Sphere},
  journal   = {Journal of Computational Physics},
  year      = {2020},
  month     = apr,
  volume    = {407},
  pages     = {109210},
  issn      = {0021-9991},
  publisher = {Elsevier BV},
  doi       = {10.1016/j.jcp.2019.109210},
  url       = {http://dx.doi.org/10.1016/j.jcp.2019.109210}
}

@article{mgritsh,
  author   = {Caldas Steinstraesser, Jo{\~a}o Guilherme and Peixoto, Pedro da Silva and Schreiber, Martin},
  title    = {Parallel-in-Time Integration of the Shallow Water Equations on the Rotating Sphere Using {Parareal} and {MGRIT}},
  journal  = {Journal of Computational Physics},
  year     = {2024},
  volume   = {496},
  pages    = {112591},
  issn     = {0021-9991},
  doi      = {10.1016/j.jcp.2023.112591},
  url      = {https://www.sciencedirect.com/science/article/pii/S0021999123006861}
}

@article{rexish,
  author   = {Schreiber, Martin and Schaeffer, Nathana{\"e}l and Loft, Richard},
  title    = {Exponential Integrators with Parallel-in-Time Rational Approximations for the Shallow-Water Equations on the Rotating Sphere},
  journal  = {Parallel Computing},
  year     = {2019},
  volume   = {85},
  pages    = {56--65},
  issn     = {0167-8191},
  doi      = {10.1016/j.parco.2019.01.005},
  url      = {https://www.sciencedirect.com/science/article/pii/S0167819118300620}
}

@article{paradiag,
  author  = {Hope-Collins, J. and Hamdan, A. and Bauer, W. and Mitchell, L. and Cotter, C.},
  title   = {{asQ}: Parallel-in-Time Finite Element Simulations Using {ParaDiag} for Geoscientific Models and Beyond},
  journal = {Geoscientific Model Development},
  year    = {2025},
  volume  = {18},
  number  = {14},
  pages   = {4535--4569},
  doi     = {10.5194/gmd-18-4535-2025},
  url     = {https://gmd.copernicus.org/articles/18/4535/2025/}
}

@article{parareal,
  author  = {Lions, J.-L. and Maday, Yvon and Turinici, Gabriel},
  title   = {A ``Parareal'' in Time Discretization of {PDE}s},
  journal = {Comptes Rendus de l'Acad{\'e}mie des Sciences - Series I - Mathematics},
  year    = {2001},
  volume  = {332},
  pages   = {661--668},
  doi     = {10.1016/S0764-4442(00)01793-6},
  url     = {http://dx.doi.org/10.1016/S0764-4442(00)01793-6}
}

@inproceedings{mgrit,
  author    = {Friedhoff, S. and Falgout, R.~D. and Kolev, T.~V. and MacLachlan, Scott P. and Schroder, Jacob B.},
  title     = {A Multigrid-in-Time Algorithm for Solving Evolution Equations in Parallel},
  booktitle = {Sixteenth Copper Mountain Conference on Multigrid Methods},
  year      = {2013},
  address   = {Copper Mountain, CO, United States},
  url       = {http://www.osti.gov/scitech/servlets/purl/1073108}
}

@article{rexi,
  author   = {Haut, T. S. and Babb, T. and Martinsson, P. G. and Wingate, B. A.},
  title    = {A High-Order Time-Parallel Scheme for Solving Wave Propagation Problems via the Direct Construction of an Approximate Time-Evolution Operator},
  journal  = {IMA Journal of Numerical Analysis},
  year     = {2016},
  month    = apr,
  volume   = {36},
  number   = {2},
  pages    = {688--716},
  issn     = {0272-4979},
  doi      = {10.1093/imanum/drv021},
  url      = {https://doi.org/10.1093/imanum/drv021},
  eprint   = {https://academic.oup.com/imajna/article-pdf/36/2/688/6767994/drv021.pdf}
}

@article{pfasst,
  author  = {Emmett, Matthew and Minion, Michael L.},
  title   = {Toward an Efficient Parallel in Time Method for Partial Differential Equations},
  journal = {Communications in Applied Mathematics and Computational Science},
  year    = {2012},
  volume  = {7},
  pages   = {105--132},
  doi     = {10.2140/camcos.2012.7.105},
  url     = {http://dx.doi.org/10.2140/camcos.2012.7.105}
}

@article{adaptivesdc,
  author    = {Saupe, Thomas and G{\"o}tschel, Sebastian and Lunet, Thibaut and Ruprecht, Daniel and Speck, Robert},
  title     = {Adaptive Time Step Selection for Spectral Deferred Correction},
  journal   = {Numerical Algorithms},
  year      = {2024},
  month     = oct,
  volume    = {100},
  number    = {1},
  pages     = {369--393},
  issn      = {1572-9265},
  publisher = {Springer Science and Business Media LLC},
  doi       = {10.1007/s11075-024-01964-z},
  url       = {http://dx.doi.org/10.1007/s11075-024-01964-z}
}

@article{adaptivepararealmd,
  author  = {Legoll, Fr{\'e}d{\'e}ric and Leli{\`e}vre, Tony and Sharma, Upanshu},
  title   = {An Adaptive Parareal Algorithm: Application to the Simulation of Molecular Dynamics Trajectories},
  journal = {SIAM Journal on Scientific Computing},
  year    = {2022},
  volume  = {44},
  number  = {1},
  pages   = {B146--B176},
  doi     = {10.1137/21M1412979},
  url     = {https://doi.org/10.1137/21M1412979}
}

@article{adaptiveparareal,
  author   = {Maday, Y. and Mula, O.},
  title    = {An Adaptive Parareal Algorithm},
  journal  = {Journal of Computational and Applied Mathematics},
  year     = {2020},
  volume   = {377},
  pages    = {112915},
  issn     = {0377-0427},
  doi      = {10.1016/j.cam.2020.112915},
  url      = {https://www.sciencedirect.com/science/article/pii/S0377042720302065}
}

@article{adaptivemgrit,
  author  = {Howse, Alexander J. and De Sterck, Hans and Falgout, Robert D. and MacLachlan, Scott and Schroder, Jacob},
  title   = {Parallel-in-Time Multigrid with Adaptive Spatial Coarsening for the Linear Advection and Inviscid Burgers Equations},
  journal = {SIAM Journal on Scientific Computing},
  year    = {2019},
  volume  = {41},
  number  = {1},
  pages   = {A538--A565},
  doi     = {10.1137/17M1144982},
  url     = {https://doi.org/10.1137/17M1144982}
}

@inproceedings{drmheat,
  author    = {et al, Huber},
  title     = {Dynamic Resource Management in {HPC} Systems Using Dynamic Processes with {PSets}},
  booktitle = {2025 IEEE 32nd International Conference on High Performance Computing, Data, and Analytics (HiPC)},
  year      = {2025},
  publisher = {IEEE},
  address   = {Hyderabad, India},
  pages     = {279--289},
  doi       = {10.1109/HiPC66333.2025.00036}
}

@book{fas,
  author    = {Briggs, William and Henson, Van and McCormick, Steve},
  title     = {A Multigrid Tutorial},
  edition   = {2},
  year      = {2000},
  publisher = {Society for Industrial and Applied Mathematics},
  address   = {Philadelphia, PA, USA},
  isbn      = {978-0-89871-462-3}
}

@inbook{defectcorrect,
  author    = {B{\"o}hmer, K. and Hemker, P. W. and Stetter, H. J.},
  title     = {The Defect Correction Approach},
  booktitle = {Defect Correction Methods: Theory and Applications},
  year      = {1984},
  publisher = {Springer Vienna},
  address   = {Vienna},
  pages     = {1--32},
  isbn      = {978-3-7091-7023-6},
  doi       = {10.1007/978-3-7091-7023-6_1},
  url       = {https://doi.org/10.1007/978-3-7091-7023-6_1}
}

@article{deferredcorrect,
  author  = {Daniel, James and Pereyra, Victor and Schumaker, Larry},
  title   = {Iterated Deferred Corrections for Initial Value Problems},
  journal = {Acta Cientifica Venezolana},
  year    = {1967},
  month   = aug,
  volume  = {19},
  pages   = {28}
}

@article{IteratedDC,
  author  = {Pereyra, V{\'i}ctor},
  title   = {Iterated Deferred Corrections for Nonlinear Operator Equations},
  journal = {Numerische Mathematik},
  year    = {1967},
  volume  = {10},
  pages   = {316--323},
  url     = {https://api.semanticscholar.org/CorpusID:123208828}
}

@InProceedings{50Years,
author="Gander, Martin J.",
editor="Carraro, Thomas
and Geiger, Michael
and K{\"o}rkel, Stefan
and Rannacher, Rolf",
title="50 Years of Time Parallel Time Integration",
booktitle="Multiple Shooting and Time Domain Decomposition Methods",
year="2015",
publisher="Springer International Publishing",
address="Cham",
pages="69--113",
isbn="978-3-319-23321-5"
}

@article{pintApps,
  author  = {Ong, Benjamin W. and Schroder, Jacob B.},
  title   = {Applications of time parallelization},
  journal = {Computing and Visualization in Science},
  year    = {2020},
  volume  = {23},
  number  = {1},
  pages   = {11},
  issn    = {1433-0369},
  doi     = {10.1007/s00791-020-00331-4},
  url     = {https://doi.org/10.1007/s00791-020-00331-4}
}

@article{convTheory,
  author  = {Bolten, Matthias and Moser, Dieter and Speck, Robert},
  title   = {Asymptotic convergence of the parallel full approximation scheme in space and time for linear problems},
  journal = {Numerical Linear Algebra with Applications},
  volume  = {25},
  number  = {6},
  pages   = {e2208},
  year    = {2018},
  doi     = {10.1002/nla.2208}
}

@article{domdecsurvey,
  author  = {Chan, Tony F. and Mathew, Tarek P.},
  title   = {Domain decomposition algorithms},
  journal = {Acta Numerica},
  volume  = {3},
  pages   = {61--143},
  year    = {1994},
  publisher = {Cambridge University Press}
}

@article{desterck2021,
  author  = {De Sterck, Hans and Falgout, Robert D. and Friedhoff, Stephanie and Krzysik, Oliver A. and MacLachlan, Scott P.},
  title   = {Optimizing multigrid reduction-in-time (MGRIT) and Parareal coarse-grid operators for linear advection},
  journal = {Numerical Linear Algebra with Applications},
  volume  = {28},
  number  = {4},
  pages   = {e2367},
  year    = {2021},
  doi     = {10.1002/nla.2367}
}

@article{desterck2025,
  author  = {De Sterck, Hans and Friedhoff, Stephanie and Krzysik, Oliver A. and MacLachlan, Scott P.},
  title   = {Multigrid Reduction-In-Time Convergence for Advection Problems: A Fourier Analysis Perspective},
  journal = {Numerical Linear Algebra with Applications},
  volume  = {32},
  number  = {1},
  pages   = {e2593},
  year    = {2025},
  doi     = {10.1002/nla.2593}
}

@article{ruprecht2018,
  author  = {Ruprecht, Daniel},
  title   = {Wave propagation characteristics of Parareal},
  journal = {Computing and Visualization in Science},
  volume  = {19},
  number  = {1},
  pages   = {1--17},
  year    = {2018},
  doi     = {10.1007/s00791-018-0296-z}
}

@article{schreiberloft2019,
  author  = {Schreiber, Martin and Loft, Richard},
  title   = {A Parallel Time-Integrator for Solving the Linearized Shallow Water Equations on the Rotating Sphere},
  journal = {Numerical Linear Algebra with Applications},
  volume  = {26},
  number  = {2},
  pages   = {e2220},
  year    = {2019},
  doi     = {10.1002/nla.2220}
}

@article{williamson2007,
  author  = {Williamson, David L.},
  title   = {The Evolution of Dynamical Cores for Global Atmospheric Models},
  journal = {Journal of the Meteorological Society of Japan. Ser. II},
  volume  = {85B},
  pages   = {241--269},
  year    = {2007},
  doi     = {10.2151/jmsj.85B.241}
}

@article{unifiedpint,
  author  = {Gander, Martin J. and Lunet, Thibaut and Ruprecht, Daniel and Speck, Robert},
  title   = {A Unified Analysis Framework for Iterative Parallel-in-Time Algorithms},
  journal = {SIAM Journal on Scientific Computing},
  volume  = {45},
  number  = {5},
  pages   = {A2275--A2303},
  year    = {2023},
  doi     = {10.1137/22M1487163}
}

@article{pfasster,
  author  = {Sch{\"o}bel, Ruth and Speck, Robert},
  title   = {{PFASST-ER}: Combining the Parallel Full Approximation Scheme in Space and Time with Parallelization Across the Method},
  journal = {Computing and Visualization in Science},
  volume  = {23},
  pages   = {12},
  year    = {2020},
  doi     = {10.1007/s00791-020-00330-5}
}

@inproceedings{spacetimeheat,
  author    = {Speck, Robert and Ruprecht, Daniel and Emmett, Matthew and Bolten, Matthias and Krause, Rolf},
  title     = {A Space-Time Parallel Solver for the Three-Dimensional Heat Equation},
  booktitle = {Parallel Computing: Accelerating Computational Science and Engineering (CSE)},
  series    = {Advances in Parallel Computing},
  volume    = {25},
  year      = {2014},
  pages     = {263--272},
  publisher = {IOS Press},
  address   = {Amsterdam, The Netherlands},
  doi       = {10.3233/978-1-61499-381-0-263}
}

@article{galewsky,
author = {Joseph Galewsky and Richard K. Scott and Lorenzo M. Polvani},
title = {An initial-value problem for testing numerical models of the global shallow-water equations},
journal = {Tellus A: Dynamic Meteorology and Oceanography},
volume = {56},
number = {5},
pages = {429--440},
year = {2004},
publisher = {Taylor \& Francis},
doi = {10.3402/tellusa.v56i5.14436},
URL = {https://doi.org/10.3402/tellusa.v56i5.14436},
eprint = {https://doi.org/10.3402/tellusa.v56i5.14436}
}

@article{williamson1,
title = {A standard test set for numerical approximations to the shallow water equations in spherical geometry},
journal = {Journal of Computational Physics},
volume = {102},
number = {1},
pages = {211-224},
year = {1992},
issn = {0021-9991},
doi = {https://doi.org/10.1016/S0021-9991(05)80016-6},
url = {https://www.sciencedirect.com/science/article/pii/S0021999105800166},
author = {David L. Williamson and John B. Drake and James J. Hack and Rüdiger Jakob and Paul N. Swarztrauber}
}

@Article{sweet,
AUTHOR = {Gaddameedi, K. and Hamon, F. and Huber, D. and Lunet, T. and S. Peixoto, P. and Caldas Steinstraesser, J. G. and Schreiber, M. and Sch\"uller, V.},
TITLE = {SWEET -- Shallow Water Equation Environment for Tests v1.0},
JOURNAL = {EGUsphere},
VOLUME = {2025},
YEAR = {2025},
PAGES = {1--32},
URL = {https://egusphere.copernicus.org/preprints/2025/egusphere-2025-5156/},
DOI = {10.5194/egusphere-2025-5156}
}

@ARTICLE{tarraf24,
  author={et al, Tarraf},
  journal={IEEE Transactions on Parallel and Distributed Systems}, 
  title={Malleability in Modern HPC Systems: Current Experiences, Challenges, and Future Opportunities}, 
  year={2024},
  volume={35},
  number={9},
  pages={1551-1564},
  doi={10.1109/TPDS.2024.3406764}
}

@article{iserte20,
author = {Iserte, Sergio and Rojek, Krzysztof},
title = {An study of the effect of process malleability in the energy efficiency on GPU-based clusters},
year = {2020},
issue_date = {Jan 2020},
publisher = {Kluwer Academic Publishers},
address = {USA},
volume = {76},
number = {1},
issn = {0920-8542},
url = {https://doi.org/10.1007/s11227-019-03034-x},
doi = {10.1007/s11227-019-03034-x},
journal = {J. Supercomput.},
month = jan,
pages = {255–274},
numpages = {20}
}

@misc{huber24,
      title={Design Principles of Dynamic Resource Management for High-Performance Parallel Programming Models}, 
      author={Dominik Huber and Martin Schreiber and Martin Schulz and Howard Pritchard and Daniel Holmes},
      year={2024},
      eprint={2403.17107},
      archivePrefix={arXiv},
      primaryClass={cs.DC},
      url={https://arxiv.org/abs/2403.17107}, 
}

@incollection{supermuc,
  author    = {Hayk Shoukourian and Arndt Bode and Herbert Huber and Michael Ott and Dieter Kranzlmüller},
  title     = {SuperMUC -- the First High-Temperature Direct Liquid Cooled Petascale Supercomputer Operated by LRZ},
  booktitle = {Contemporary High Performance Computing: From Petascale toward Exascale},
  volume    = {3},
  editor    = {Jeffrey S. Vetter},
  publisher = {CRC Press},
  address   = {Boca Raton, FL, USA},
  year      = {2019},
  doi       = {10.1201/9781351036863-10}
}

@online{dynres,
  author = {Dominik Huber},
  title = {DynRes Software Stack},
  year = 2026,
  url = {https://dynres.readthedocs.io/en/latest/},
  urldate = {2026-09-11}
}

@book{hairer,
author = {Hairer, Ernst and Wanner, Gerhard},
year = {1996},
month = {01},
pages = {},
title = {Solving Ordinary Differential Equations II. Stiff and Differential-Algebraic Problems},
volume = {14},
journal = {Springer Verlag Series in Comput. Math.},
doi = {10.1007/978-3-662-09947-6}
}

@InProceedings{ju25,
author="et al, Ju",
editor="Blaas-Schenner, Claudia
and Niethammer, Christoph
and Haas, Tobias",
title="Dynamic Resource Management for In-Situ Techniques Using MPI-Sessions",
booktitle="Recent Advances in the Message Passing Interface",
year="2025",
publisher="Springer Nature Switzerland",
address="Cham",
pages="105--120",
isbn="978-3-031-73370-3"
}

@article{posner25,
author = {Posner, Jonas and Ellersiek, Tim and Bietendorf, Nick and Huber, Dominik and Schreiber, Martin and Schulz, Martin},
title = {Toward Dynamic Resource Management: An Asynchronous Many-Task (AMT) Runtime System leveraging Dynamic Processes with PSets (DPP)},
year = {2025},
issue_date = {Dec 2025},
publisher = {Springer-Verlag},
address = {Berlin, Heidelberg},
volume = {6},
number = {8},
url = {https://doi.org/10.1007/s42979-025-04405-3},
doi = {10.1007/s42979-025-04405-3},
journal = {SN Comput. Sci.},
month = nov,
numpages = {16}
}

\appendix

\end{document}